\documentstyle[graphicx]{mn}

\newif\ifAMStwofonts
\AMStwofontstrue

\def\ngc{{NGC 5548}}

\def\einstein{{\it Einstein}}
\def\exosat{{\it EXOSAT}}
\def\xmm{{\it XMM-Newton}}
\def\chandra{{\it Chandra}}

\def\et{{et al.\ }}

\def\rosat{{\it ROSAT}}
\def\ginga{{\it GINGA}}
\def\euve{{\it EUVE}}
\def\heao{{\it HEAO-1} A2}
\def\asca{{\it ASCA}}
\def\sax{{\it BeppoSAX}}
\def\xte{{\it RXTE}}

\newcommand{\ls}{\mathrel{\hbox{\rlap{\hbox{\lower4pt\hbox{$\sim$}}}\hbox{$<$}}}}
\newcommand{\gs}{\mathrel{\hbox{\rlap{\hbox{\lower4pt\hbox{$\sim$}}}\hbox{$>$}}}}

\def\arcs{{\hbox{$^{\prime\prime}$}}}

\def\Msun{\hbox{$\rm ~M_{\odot}$}}

\def\atpcm{{\rm ~atoms~cm^{-2}}}

\def\H0{{\rm ~km~s^{-1}~Mpc^{-1}}}

\def\msun{M_{\rm \odot}}

\def\et{{et al.}}

\title[An \xmm\ observation of \ngc]
        {A simultaneous \xmm\ and \sax\ observation of the archetypal 
        Broad Line Seyfert 1 galaxy \ngc}
\author[K.A.Pounds \et]
        {K.A.Pounds,$^{1}$
         J.N.Reeves,$^{1}$
         K.L.Page,$^{1}$
         R.Edelson,$^{2}$
         G.Matt$^{3}$
         and G.C.Perola$^{3}$\\
$^1$ Department of Physics and Astronomy, University of Leicester,
Leicester LE1 7RH, UK\\
$^2$ Astronomy Department, UCLA, Los Angeles, CA 90095-1562, USA\\
$^3$ Dipartimento di Fisica, Universita di Roma Tre, I-00146, Roma,
Italy\\
}
\date{Accepted 2003 January 30; Submitted 2002 October 11}
\pagerange{\pageref{firstpage}--\pageref{lastpage}}
\pubyear{2002}
\begin{document}
\maketitle
\label{firstpage}

\begin{abstract}

We report the spectral analysis of a long \xmm\ observation
of the well-studied, moderate luminosity Broad Line Seyfert 1 galaxy \ngc.
The source was at an historically average brightness and we find the hard 
(3-10~keV) spectrum can be well fitted by a power law
of photon index $\Gamma$$\sim$1.75, together with reflection.
The only feature in the hard X-ray spectrum is a
narrow emission line near 6.4~keV, with an equivalent width of
$\sim$60 eV.
The energy and strength of this line is consistent with
fluorescence from `neutral' iron distant from the central continuum
source. We find no
evidence for a broad Fe K line, with an upper limit well below previous
reports, suggesting the inner accretion disc is now absent or
highly ionised.
The addition of simultaneous \sax\ data allows the analysis to be
extended to 200 keV, yielding important constraints on the total reflection.
Extrapolation of the hard X-ray power law down to 0.3 keV shows a
clear `soft excess' below $\sim$0.7 keV.  
After due allowance for the effects of
a complex warm absorber, measured with the \xmm\ RGS, we
find the soft excess is better described as a smooth upward curvature 
in the continuum
flux below $\sim$2~keV.
The soft excess can be modelled either by Comptonised thermal
emission or by enhanced reflection from the surface of a highly ionised disc.
\end{abstract}


\section{Introduction}

It is widely believed that the primary luminosity of an AGN originates
in an accretion disc around a super-massive black hole. However, in the X-ray band,
a hard power-law component generally dominates above $\sim$2~keV in the 
well studied Broad Line Seyfert~1 galaxies and is considered to arise in a hot corona 
above the surface of the accretion disc, where
optical/UV photons from the disc are Comptonised to X-ray energies.
These X-rays in turn illuminate the disc, being either `reflected'
towards the observer or thermalised back into
optical/UV emission (Mushotzky \et\ 1993). In the X-ray spectral band,
evidence for this disc reflection component, in the form of a 
fluorescent Fe K$\alpha$ line
near 6.4~keV, an Fe K edge at $>7$~keV and a
Compton `hump' at $>10$~keV, is found in many 
bright Seyfert~1 galaxies (e.g. Pounds \et\ 1990, Nandra \& Pounds 1994).
The improved resolution of \asca\ has more recently been used to show 
that the Fe K$\alpha$ line is
often broad, with excess flux, particularly in a `red' wing to the
line (Tanaka \et\ 1995, Nandra \et\ 1997), believed due to emission from the inner disc in a region of both high
velocities and high gravity (Fabian \et\ 2000).

The substantial gains in sensitivity, bandwidth and spectral
resolution of \xmm\
and \chandra\ are now further qualifying this standard
picture,
with indications that the energy,
breadth and strength of the broad Fe K$\alpha$ line (the inner disc
component) is critically dependent on how the underlying X-ray continuum
is modelled. Also, the frequent
detection of a narrow emission line at $\sim$6.4~keV,
probably
having a distinct origin, away from the inner disc
(Reeves \et\ 2001, Yaqoob \et\ 2001, Kaspi \et\ 2001, Matt \et\ 2001,
Gondoin \et\ 2001, Pounds \et\ 2001), now has to be
accounted for in determining the profile of any broad emission line
component (Pounds and Reeves 2002, Weaver and Reynolds 1998).

In this paper we report on a simultaneous \xmm\ and \sax\ observation of \ngc, 
one of the brightest and
best-studied, medium luminosity Broad Line Seyfert 1 galaxies.
At $z=0.017$ (with $ H_0 = 75 $~km\,s$^{-1}$\,Mpc$^{-1})$ \ngc\ has 
an X-ray luminosity (2--10~keV) which historically lies in the range
$1-4\times10^{43}$~erg s$^{-1}$.
The Galactic absorption column towards \ngc\ is $1.65\times10^{20}$~cm$^{-2}$
(Murphy \et\ 1996), rendering it easily visible over the whole
($\sim$0.2--12~keV) spectral band of the EPIC and RGS detectors on \xmm.

The hard X-ray spectrum of \ngc\ obtained with the
\ginga\ satellite (Nandra and Pounds 1994) showed the 2--18 keV
spectrum to be well described by a power law
of index $\Gamma$$\sim$1.65-1.85, including a reflection factor
R~=~$\Omega$/2$\pi$, where $\Omega$ is the solid angle subtended by
the reflecting matter, in the range
0.5--1, with an Fe K emission line (equivalent width $\sim$110~eV at
$\sim$6.4~keV) and the imprint of low energy absorption by ionised gas 
of equivalent hydrogen column density $5\times10^{22}$ $\atpcm$.
Fitting a narrow line to an early \asca\ observation refined
the line energy to $E=6.39\pm0.04$~keV, with an EW of $\sim$90~eV,
increasing to $\sim$170~eV for a broad line fit (Nandra \et\ 1997a).

Evidence for a `soft excess', whereby
the X-ray spectrum below $\sim$1~keV lies above an extrapolation of 
the hard X-ray
power law, was first implied in spectra of several AGN obtained by 
\heao\ (Pravdo \et\
1981), \exosat\ (Arnaud \et\
1985; Turner \& Pounds 1989) and \einstein\ (Bechtold \et\ 1987).
However, quantifying the soft excess has remained uncertain, since its
determination depends on knowledge of both the power law
component and the effects of overlying absorption. 
The physical origin of any soft excess is also generally unclear, since thermal 
emission from the
accretion disc around a $10^{6}$-$10^{8}$ $\msun$ black hole (e.g. Arnaud 
\et\ 1985; 
Czerny \& Elvis 1987), should be much cooler than the $\sim$100~eV
soft X-ray temperatures typically inferred, without
secondary heating or reprocessing of the disc photons.
Understanding the origin of the disc emission, and its interaction
with the `corona' nevertheless remains a key to understanding the whole
X-ray emission mechanism in Seyfert galaxies.

In the case of \ngc\ an early \exosat\ 
observation
(Branduardi-Raymont 1986) found evidence for a soft excess emerging
when the 2-6 keV flux was above some average level. 
This conclusion gained support from an extended \rosat\ observation 
(Done \et 1995) where the 0.2-0.7 keV flux from \ngc\ was found to
vary more strongly than that at 1-2 keV.
However, in a long \sax\ observation in 1997, Nicastro \et\
(2000) found the spectral variability in \ngc\ was primarily
due to a change in the power law slope, with no requirement for a soft
excess. Other results of this \sax\ study were: a high energy cut-off in
the range 90--165 keV; low energy absorption well-modelled by 
OVII and OVIII edges; 
an unresolved Fe K emission line and a continuum `hump' 
around 20--30 keV consistent with reflection R $\sim$0.5 from `cold'
matter.

\section{Observation and data reduction}

\ngc\ was observed on 2001 July 9/10 and 12 during orbits 290 and 291 in the
\xmm\ Guest Observer programme. The observations with the EPIC MOS2
(Turner \et 2001) and PN
(Str\"{u}der \et 2001) detectors were
of order $\sim$90~ksec in orbit 290 and $\sim$35~ksec in orbit 291,
with medium and thin filters, respectively. MOS1 data were in
timing mode and therefore not useful in the present analysis. The 
simultaneous RGS observations were of $\sim$90~ksec and $\sim$40~ksec,
during orbits 290 and 291, while the \sax\ observation covered the
period 8--10 July, overlapping the \xmm\ longer (orbit 290) exposure.

\sax\ data were reduced in the standard way (Guainazzi \et 1999). MECS
counts were extracted from a region of 4 arc min. radius, and the PDS
spectrum was extracted using the fixed pulse rise-time threshold.

The EPIC data were first screened with the XMM SAS v5.3 software.
X-ray events corresponding to patterns 0-12 for the MOS2 camera
(similar to grades 0-6
in \asca) were used; for the PN 
patterns 0-4 (single and double pixel events) were selected.
A low energy cut of 200 eV was applied to all data
and known hot or bad pixels were removed.
The non X-ray background remained low until the last $\sim$5 ksec of
the orbit 290 observation, which were excluded from the subsequent
spectral analysis.

We extracted source and background spectra for
the PN and MOS2
detectors with a circular source region of
45\arcs\ radius defined around the
centroid position of \ngc, with the background being taken from an
offset position close to the source.
The 2-10 keV X-ray flux from \ngc\ varied between
$5-7\times10^{-11}$~ergs~cm$^{-2}$~s$^{-1}$ during the \xmm\
observations, lying at the upper end (though near the mean) of the 
historical range; however, 
photon pile-up was negligible in the Small Window mode
chosen for all EPIC observations.
The individual spectra 
were binned to a minimum of 20 counts per
bin, to facilitate use of the $\chi^2$ minimalisation technique in
spectral fitting. 

The primary analysis in this paper is based on observations from the 
first \xmm\
orbit, which has the longest exposure and also benefits from
simultaneous \sax\ data. In the
present analysis we use both 
MOS2 and PN data to evaluate the
hard X-ray spectrum, but only the MOS2 data to study the `soft excess', 
since the
MOS2 broad band calibration is considered to be more certain than that of the 
PN at present. Response functions for spectral fitting to the \xmm\
data were generated from the SAS v5.3. We use the MECS and PDS data from \sax\ to constrain
the continuum reflection and high energy cut-off. 

All spectral fits include absorption due to the line-of-sight
Galactic column of $N_{H}=1.65\times10^{20}\rm{cm}^{-2}$. 
All fit parameters 
are given in the AGN rest-frame.
Errors are quoted at
the 90\% confidence level (e.g. $\Delta \chi^{2}=2.7$ for one
interesting parameter).

\section{The hard power law spectrum}  

It has long been clear that the key to modelling the X-ray spectra of
AGN is to first
identify, if possible, the `underlying continuum' on which additional
emission and absorption features are superimposed.
We therefore began our analysis of \ngc\ by fitting the \xmm\ data above 3 keV, where any
effects of intervening absorption should be negligible, a view
supported by the RGS analysis of the warm absorber (Steenbrugge \et\
2003). A power law gave a good fit to both MOS2 and PN data over the
3-10 keV band apart from a clear 
excess flux
near 6.4 keV. The respective parameters are listed in fit 1 of
Tables 1 and 2. The power law index was flatter by $\sim$ 0.02
in the PN fit, a difference that is now only marginally significant and
suggests an encouraging convergence in the latest calibrations of the EPIC 
cameras.  Figure 1 reproduces the data:model ratio of the MOS2 and PN power law
fits. No other features are seen in addition to the narrow emission line at
$\sim$ 6.4 keV.

\subsection{The Fe K band}
The addition of    
a narrow Gaussian emission line to the simple power law fits,    
with energy, width and flux free, produced a very significant improvement
to both MOS2 and PN spectral fits. These improved fits are detailed in
fit 2 of Tables 1 and 2, respectively. Inclusion of the emission line in the
3--10 keV fits increased the power law 
index by                                                            
$\sim$0.03 for each data set.                                          
The line is narrow and
lies at a rest energy consistent with fluorescence from neutral iron.
More specifically, the upper limit of 6.41 keV in both MOS2 and PN fits,
corresponds to ion stages between FeI and FeX, while it is interesting to
note that the high statistical significance of the EPIC data
gives a line width with similar formal precision to that reported from
an earlier observation with the \chandra\ HETG (Yaqoob \et\ 2001). We
return to this point in the Discussion (section 5). 

The formal 3 $\sigma$ upper limit for an additional 
broad line, modelled as a `diskline' (Fabian \et\ 1989) with
energy and profile as reported from an \asca\ observation
(Nandra \et\ 1997) is 43 eV. In contrast, the \asca\ analysis, which
recorded a power law index and flux almost identical to the present
observation, found a broad line equivalent width of 200$\pm$90 eV.

\begin{figure}
\centering
\includegraphics[width=5.8 cm, angle=270]{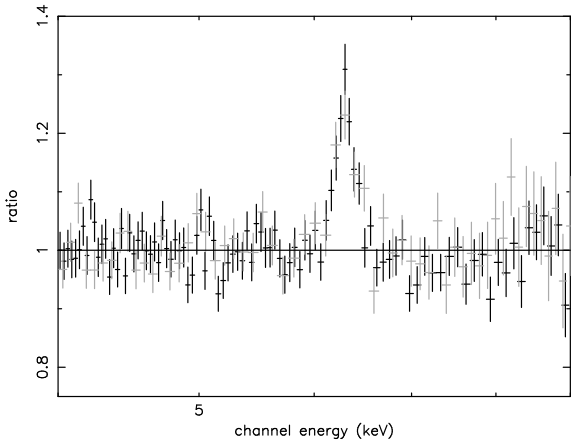}
\caption
{MOS2 (black) and PN (grey) spectral fits over the 3--10 keV band shown 
as the ratio
of the data to the best power law fit in each case. The only spectral
feature evident is a narrow emission line at a rest energy of $\sim$
6.4 keV.}
\end{figure}

\begin{figure}                                                          
\centering                                                              
\includegraphics[width=5.5 cm, angle=270]{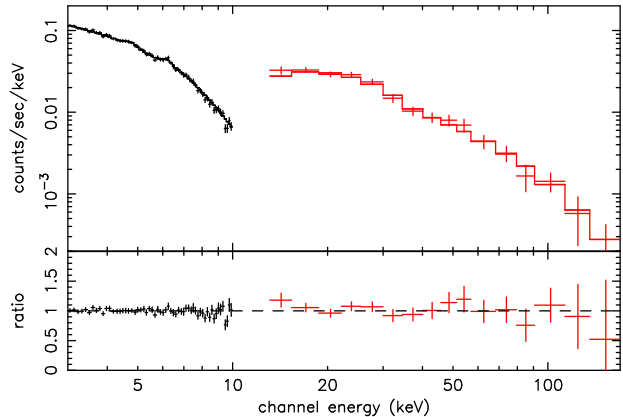}                     
\caption                                                                
{Combined MECS and PDS spectral fit over the 3--200 keV range compared  
to the cutoff power law plus cold reflection model of Table 1, fit 3.   
The upper panel shows the count spectra                                 
(crosses) and the model folded through the detector                     
responses (histograms).  The lower panel shows the fit residuals.}      
\end{figure}                                                                                               
                                       
\subsection{Cold reflection}
An important limitation in \xmm\ data is a rapid loss of sensitivity
above $\sim$10 keV, preventing the unambiguous determination of continuum
reflection, which is in general required to determine the true slope of
the primary (power law) continuum.
Fortunately we have
simultaneous \sax\ data on \ngc\ for the longer, orbit 290 observation, 
which extends the spectral
coverage to $\sim$200 keV and constrains the continuum `hump' caused
by down-scattering of high energy photons and is a key measure of
reflection.
We first compared the MOS2, PN and MECS data, finding good agreement in
describing the \ngc\ spectrum over the band 3--10 keV.
The PDS and MECS data were then combined, and tested against a model
with components of a cut-off power law, a narrow Fe K line and cold
reflection
modelled by PEXRAV. All fit parameters were tied for the 2 detectors,
with an additional normalisation constant of 0.86 for the PDS (Fiore
\et 1999).
An excellent fit (figure 2) yielded a cold reflection factor
R $\sim$0.7 which, in turn, led to
a steepening of the powerlaw
index by  $\sim$0.05. Addition of the MOS2 data, with again only the
normalisation untied, confirmed the excellence of this fit, 
details of which are listed in
Table 1, fit 3. The parameters of the reflection fit were found to be
only a weak function of the assumed iron abundance. In the best fit, quoted,
the abundance was left free and determined to be $1.4\pm0.6$ solar.
The cosine
of the disc inclination was fixed at 0.9. 
The PN, MECS and PDS data were next fitted with the above model,
yielding the parameters detailed in Table 2, fit 3. 

In summary, an overall and self-consistent description of the
3--200 keV spectrum consists of a cut-off power law and cold
reflection
at a level that could also explain the narrow Fe K emission line (not
included in PEXRAV).
The observed 3--100 keV flux was $1.2\times10^{-10}$~ergs~cm
$^{-2}$~s$^{-1}$.

\begin{table*}
\centering
\caption{Spectral fits to joint
MOS, MECS and PDS data over 3--10 keV and 3--200 keV energy bands.
$^a$ Rest energy of line (keV).
$^b$ Intrinsic (1 sigma) width of line (eV)
$^c$ Equivalent width of line (eV)
$^{d}$ Energy cut-off (keV)
$^{e}$ Reflection parameter of cold matter; R~=~$\Omega$/2$\pi$}

\begin{tabular}{@{}lcccccccc@{}}
\hline
Fit & $\Gamma$ & \multicolumn{3}{c} {Fe K line} & \multicolumn{2}{c}
{Reflection component} & $\chi^{2}$/dof \\

\ & \ & E$^a$ & $\sigma^b$ & EW$^c$ & E$_{c}^{d}$ & R$^{e}$\\

\hline

1. P-L only & 1.67$\pm$0.02 & & & & & & 426/399 \\

2. P-L + Gauss & 1.70$\pm$0.02 & 6.39$\pm$0.02 & 40$\pm$40 &
63$\pm$18 & & & 354/396 \\

3. P-L + Gauss + Pexrav & 1.75$\pm$0.02 & 6.39$\pm$0.02 & 45$\pm$45 &
63$\pm$18 &240$\pm$90 &0.75$\pm$0.19 &408/467 \\
\hline
\end{tabular}
\end{table*}

\begin{table*}                                                        
\centering                                                            
\caption{Spectral fits to joint                                       
PN, MECS and PDS data over 3--10 keV and 3--200 keV energy bands.    
$^a$ Rest energy of line (keV).                                       
$^b$ Intrinsic (1 sigma) width of line (eV)                           
$^c$ Equivalent width of line (eV)                                    
$^{d}$ Energy cut-off (keV)                                           
$^{e}$ Reflection parameter of cold matter; R~=~$\Omega$/2$\pi$}

\begin{tabular}{@{}lcccccccc@{}}                                      
\hline                                                                
Fit & $\Gamma$ & \multicolumn{3}{c} {Fe K line} & \multicolumn{2}{c}  
{Reflection component} & $\chi^{2}$/dof \\

\ & \ & E$^a$ & $\sigma^b$ & EW$^c$ & E$_{c}^{d}$ & R$^{e}$\\         

\hline

1. P-L only & 1.65$\pm$0.01 & & & & & & 1511/1353 \\

2. P-L + Gauss & 1.68$\pm$0.01 & 6.40$\pm$0.01 & 64$\pm$24 &          
67$\pm$14 & & & 1347/1350 \\

3. P-L + Gauss + Pexrav &
 1.73$\pm$0.02 & 6.39$\pm$0.02 & 60$\pm$24 & 
60$\pm$18 &275$\pm$100 &0.58$\pm$0.15 &1440/1465 \\                      
\hline                                                                
\end{tabular}                                                         
\end{table*}

\begin{table*}
\centering
\caption{Broad-band (0.3--10 keV) continuum spectral fits to MOS2 data.
$^a$~Photon Index of power-law.
$^b$~Temperature (eV) of blackbody emission or Comptonising
electrons.
$^{c}$~Optical depth of Comptonising electron distribution.
$^{d}$~Ionisation parameter.
$^{e}$~DB = Fit to ionised reflection model (Ballantyne \et\ 2002) - see
text for details. }

\begin{tabular}{@{}lccccccc@{}}
\hline
Fit & $\Gamma^{a}$ & $kT_{1}^{b}$ & $kT_{2}^{b}$ 
& $\tau^{c}$ & xi$^{d}$ &
$\chi^{2}$/dof\\
\hline

1. PL+2$\times$bbody & 1.76$\pm$0.03 & 130$\pm$8 & 365$\pm$19  & 
& & 551/566 \\
2. PL+compTT & 1.75$\pm$0.03 & & 400$\pm$60 & 16$\pm$4 & &
581/569 \\
3. PL+DB$^{f}$ & 1.75$\pm$0.02 & & & & 4.1$\pm$0.2
& 586/569\\
\hline
\end{tabular}
\end{table*}

\section{The Soft Excess}

Extrapolating the best-fit description of the hard X-ray spectrum 
(fit 3 in Table 1)
to the lower well-calibrated limit of the EPIC MOS camera ($\sim$~0.3 keV)
shows an apparently unambiguous excess of soft X-ray flux below 
$\sim0.7$~keV (figure 3).
There is also clear evidence for absorption which `cuts' a significant 
fraction from any smooth profile continuum between $\sim$~0.7-2 keV,
a factor that appears to be common in the EPIC spectra of low and medium
luminosity AGN and must be accounted for before
attempting to model any soft excess emission (Pounds and
Reeves 2002). 
In the present case the simultaneous \xmm\ RGS data
(Steenbrugge \et\ 2003), confirm the presence of complex absorption
and associate it with an ionised outflow. We have worked with colleagues at 
SRON to assess joint
fits to the EPIC and RGS data for \ngc\ (see Figure 4). Our conclusions, 
are that the data are in good overall agreement,
though the RGS data are relatively insensitive to the precise
continuum shape, which in the joint fit is effectively determined by the
EPIC data. For the present EPIC analysis a simpler description of
the warm absorber was used, with advantages in terms of 
modelling speed and in being a better match to the resolution of the
EPIC detectors. This working description of the warm absorber is in
terms of a series of absorption edges at (rest frame) $\sim$0.725 keV ($\tau$
$\sim$0.29) ,
$\sim$0.87 keV ($\tau$$\sim$0.28), $\sim$1.23 keV ($\tau$$\sim$0.08), 
and $\sim$1.78 keV ($\tau$$\sim$0.06). The energy and optical depth
of these edges were left free in fitting to the
EPIC data, and correspond
approximately to the absorption edges of O VII, O VIII, Ne
IX/X
and Mg XI/XII. Visual comparison of figures 4 and 5 indicates why the 
absorption edge set is a reasonable substitute for the complex RGS absorption
spectrum at the resolution level of the EPIC detectors.
We note the two deepest edges,
are consistent with the main
features found
in \rosat\ and \asca\ modelling, and account for the strongest
absorption,
from O VII, O VIII and iron M- and L-shell ions, while Ne K-shell
absorption is also clearly seen.

Fixing the cold reflection factor and Fe K line parameters as
determined from the 3-200 keV fit,
we then
proceeded to model the `soft excess'. A second power law, or a single
blackbody,
gave a very poor fit, the breadth of the soft excess
requiring 2 blackbody components to obtain an acceptable value of
chi-squared.
Table 3 (fit 1) details this best-fit to the overall (0.3-10 keV)
spectrum, which is reproduced in Figure 5. 
Parameterised in this way, the soft excess is found to be
energetically a minor part of the X-ray flux from \ngc, the multiple
blackbody component accounting for only
$\sim$20\% of the observed
0.3--1 keV flux of
$2.1\times10^{-11}$~erg cm$^{-2}$ s$^{-1}$,
$\sim$12\% of the 1--2 keV                                            
flux of                                              
$1.5\times10^{-11}$~erg cm$^{-2}$ s$^{-1}$, and a
negligible fraction of the harder (2--10keV)
flux of                                             
$5.0\times10^{-11}$~erg cm$^{-2}$ s$^{-1}$.

\begin{figure}
\centering
\includegraphics[width=5.5 cm, angle=270]{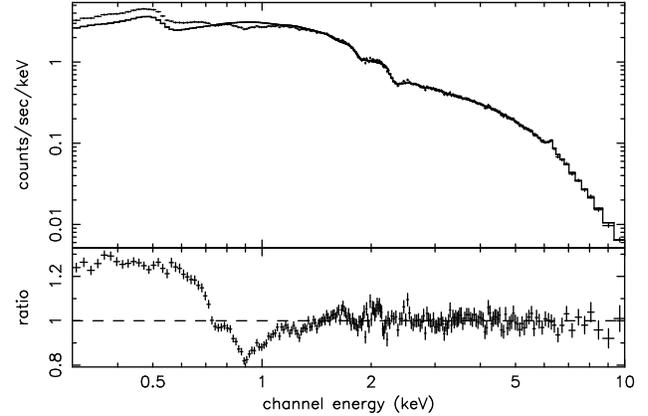}
\caption
{Hard X-ray spectral fit above 3 keV, fit 3 of Table 1, extrapolated to 0.3 keV showing
the 
soft excess modified by line-of sight absorption (see text for
details).}
\end{figure}

\begin{figure}                                                      
\centering                                                          
\includegraphics[width=6.3 cm, angle=270]{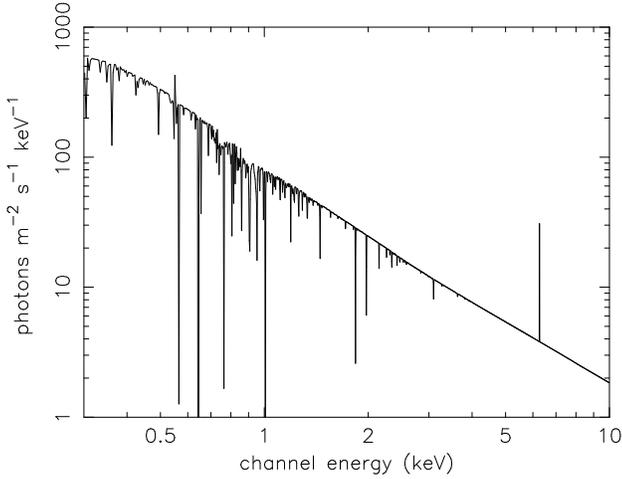}                 
\caption                                                            
{Joint MOS2 and RGS data fit for \ngc\ over the 0.3--10 keV band.  
The underlying continuum is modelled with a power law and two
blackbodies (as in Table 3, fit 1), with complex absorption determined by 
the RGS.}                                                      
\end{figure}

\begin{figure}                                                      
\centering                                                          
\includegraphics[width=6 cm, angle=270]{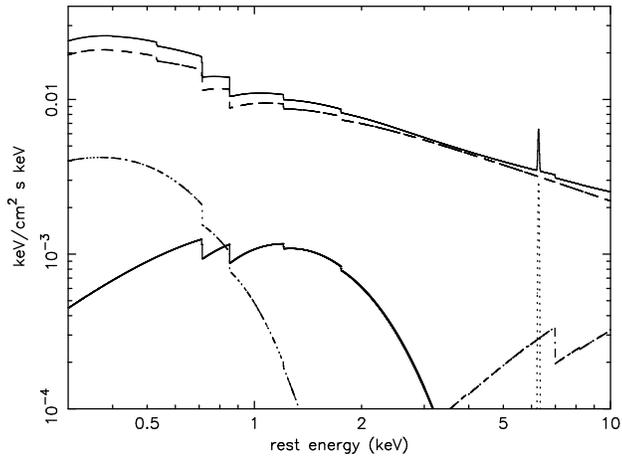}                
\caption                                                            
{Model fit to the MOS2 data over the 0.3--10 keV range              
with a power law plus 2 blackbody emitters as detailed in Table 3,  
fit 1.                                                              
The warm absorber is modelled here with a set of 4 absorption edges   
as described in the text.}                                          
\end{figure}

\subsection{Comptonised disc emission}

While the addition of multiple blackbody components is a useful way 
to parameterise the
soft excess it reveals little about the physical origin of this
emission component, since - as noted earlier - the intrinsic thermal radiation
of an accretion disc around a supermassive black hole is expected to be
much cooler than implied by the observed soft X-ray spectrum. 

To attempt a more physical description of the spectral curvature
underlying the complex absorption, we therefore replaced the blackbody
components with Comptonised thermal disc   
emission, a model successfully used to describe \xmm\ detections of 
strong soft 
excesses in PKS 0558-504 (O'Brien \et\ 2001) and 1H 0419-577 (Page \et\   
2002), and applied to \sax\ data on \ngc\ (Petrucci \et\ 2000).                                                                  
Specifically, we used the compTT model in xspec which gave a good fit to the 0.3--10 keV MOS 
spectrum of \ngc, 
with parameters as detailed in Table 3, fit 2 and illustrated in
Figure 6. We assumed an input
photon temperature of 25 eV, and found an optically thick scattering 
region ($\tau$$\sim$16) of
kT $\sim$0.4 keV. These parameters were found to be relatively insensitive 
to the
input photon temperature over a factor of 2-3. Since they are also known 
to be strongly
covariant we show in Figure 7 the range of acceptable values for the
temperature and optical depth. We interpret the 
relatively low temperature of the scattering medium
(compared with previous fits to the soft excess of PKS 0558-504 and 1H
0419-577) in terms of matching the less broad (less hot) soft 
is excess in \ngc. An important prediction of this fit is that soft
X-ray variability would be reduced by the large optical depth (in
contrast to the conclusions of Done \et 1995).                     

\subsection{Ionised reflection}

We next attempted a fit 
with an ionised reflector model (eg Ballantyne \et\
2001). In this model when the disc surface is       
highly ionised it produces a smooth excess of emission in the 
soft band, with relatively weak spectral features.
The result was again a good fit to the soft X-ray excess, with a
ionised reflection factor of R $\sim$0.6. Other 
details of this fit are given in Table 3, fit 3, and the model is
reproduced in figure 8.

At the high ionisation parameter implied by this fit, individual 
spectral features are smeared out, due to the occurrence
of multiple ion stages and to the effects of electron scattering in 
the upper layers of the reflector. 
This smearing is critical in fitting to our \ngc\ observation over the 
$\sim$~ 7--9 keV band where deep absorption edges of ionised Fe would
otherwise be in conflict with the data. 

\begin{figure}  
\centering  
\includegraphics[width=6 cm, angle=270]{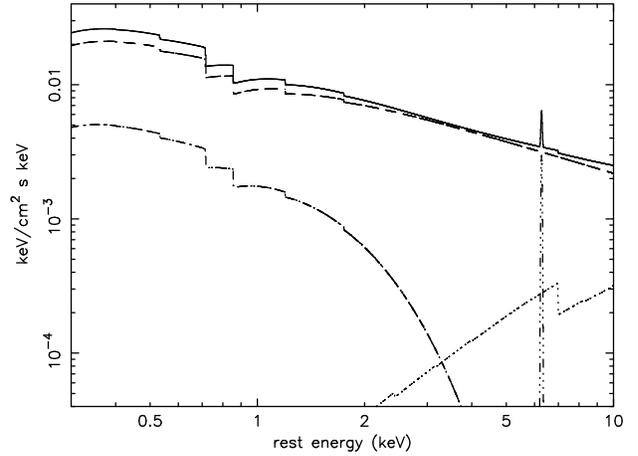}  
\caption  
{Model fit to the MOS2 data over the 0.3--10 keV range  
with a power law plus Comptonised thermal emission as detailed in 
Table 3, fit 2. The warm absorber is again modelled with a set of
4 absorption edges.}  
\end{figure}  

\begin{figure}                                                      
\centering                                                          
\includegraphics[width=6 cm, angle=270]{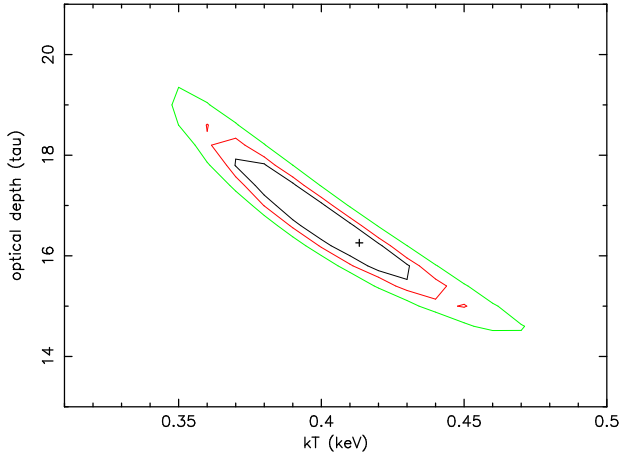}                
\caption                                                            
{1,2 and 3 sigma contours of the covariant parameters of temperature and optical
depth in the Comptonised model fit to the soft excess.}             
\end{figure}

\begin{figure}                                                     
\centering                                                         
\includegraphics[width=6 cm, angle=270]{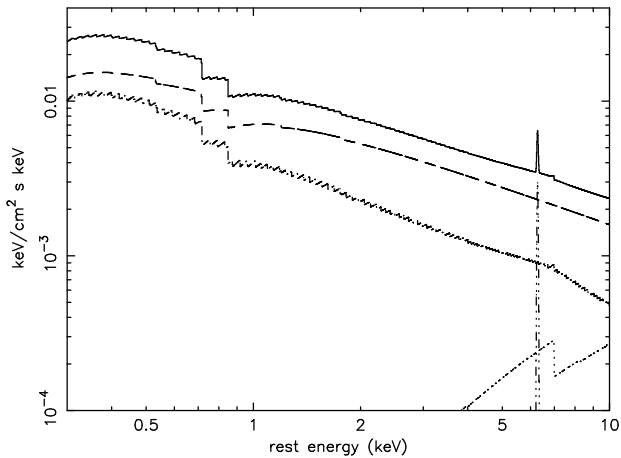}                   
\caption                                                           
{Model fit to the MOS2 data over the 0.3--10 keV range                      
with a power law plus ionised reflection (lower curve) as detailed in Table 3, line
3. The warm absorber is again modelled with a set of 4 absorption edges.} 
\end{figure}                                                         

\section{Discussion}

Our analysis emphasises the importance of simultaneous broad-band data to
de-convolve the
several inter-related components in the observed X-ray spectrum of AGN.
Without such data, progress in understanding the primary emission and
re-processing mechanisms will remain elusive. The key importance of
measuring continuum reflection is underlined in our study of \ngc\ where it has
a significant effect on the deduced hard X-ray power law slope, which
in turn is
critical in quantifying any soft excess component. The impact of
over-lying absorption on the emerging
emission in the soft X-ray band must also be properly accounted for to
determine the
true soft emission spectrum.

With the above constraints in mind our analysis of the long    
observation of \ngc\ on 9-10 July 2001 proceeded by first determining
the amount of cold reflection using the simultaneous \sax\ data.    
The value obtained, R $\sim$0.58-0.75 for the \sax\ plus PN and \sax\
plus MOS2 fits, 
is compatible with the strength of the narrow Fe
K line,
from which one conclusion could be that the only significant
reflection      
is from cold matter some distance from the central continuum source.    
That would imply that the inner accretion disc is either 
absent or highly ionised, a conclusion consistent with the absence of
a broad Fe K line.    

In the latter respect it is interesting that one description of
the (relatively weak) soft excess is by enhanced reflection from    
highly ionised matter, presumably the inner accretion disc. However
this     
interpretation of the soft excess is not unique, and we have no useful 
constraints in terms of either ionised absorption edges in the    
$\sim$8-9 keV region (where the \xmm\ sensitivity is rather low)    
or a definite contribution to the continuum `hump' at higher energies.
The latter point is considered further in section 5.3.

\subsection{Narrow Fe K line}
The best prospect for clarifying the origin of the 
narrow Fe K line,
with current observational facilities, appears to be via a careful search for
variability  
with the statistical precision of \xmm\ EPIC data, or by constraining the 
line width with the HETG on
\chandra.
A comparison of our present data with the \chandra\ observation of
Yaqoob \et\ (2001) provides a basis for such a future study in the
case of \ngc.

Taking our PN data, which has the best statistics on the iron line, we
find a line energy of $6.40\pm0.01$ keV, consistent with the \chandra\ 
line energy of $6.402\pm0.026$ keV. The line fluxes are also
consistent, albeit not well constrained. Given current uncertainties
in cross-calibration between the two observations, we
note instead that the continuum flux was a factor of 2 lower during the
\chandra\ observation, while the equivalent width was double the \xmm\
value ( Yaqoob \et\ 
2000 find $133\pm58$ eV, compared with our \xmm\ figure of $60\pm18$
eV),
suggesting that the
bulk of the narrow line may have remained constant over the 17-month interval
between the \chandra\ and \xmm\ observations. 

However, we also note that the PN data marginally resolve the `narrow' line,
with $\sigma$ = $60\pm24$ eV ($6500\pm2600$ km s$^{-1}$ FWHM). The comparable 
\chandra\ HETG
parameters were 4500$^{+3560}_{-2590}$ km s$^{-1}$ FWHM. Effectively the much better
statistics of the \xmm\ detection are compensating for the fourfold worse
energy resolution, lending some support to the conclusion that the
`narrow' Fe K line in \ngc\ is marginally resolved. If confirmed, the 
implication of this result is
that a significant component of the `narrow' Fe K line arises at distances from the
hard X-ray source comparable to the Broad Line clouds.
  
\subsection{Soft Excess}
Having obtained a reliable measure of the
dominant power law component, extrapolation to the lower useful energy
limit of EPIC confirms the presence of a
`soft excess'. Given the simultaneous measure of the `warm absorber'
by the \xmm\ RGS we feel confident that - in this case - the soft
excess is reliably determined. We find it to be characterised by a
smooth upward curvature, probably typical of unobscured, radio-quiet AGN 
(Pounds
and Reeves 2002), which can be parameterised by blackbody emitters of
$\sim100$  and $\sim300$ keV.

An explanation of the soft excess remains unclear, however. Two 
physical models, which we find to be equally good fits to the EPIC  
data, have different inferences for the X-ray emission process. Ionised 
reflection is an attractive explanation, given the energetically
dominant and relatively hard power law spectral component in \ngc.
This implies that in \ngc\ - which we point out may be an archetypal 
Broad Line Seyfert 1 galaxy - the X-ray emission is primarily
due to accretion energy being deposited in a low density, high 
temperature medium (eg a `corona'),
and there is little or no intrinsic X-ray emission from the accretion disc.

On the other hand, if the soft excess arises by thermal disc emission, upscattered into the soft  
X-ray band by passage through an optically thick, hot `skin' (eg
Hubeny \et\ 2001), it
offers the important possibility of studying the thermal structure and
internal heating of the inner disc. Some support is provided for the Comptonisation model by the
finding of a time lag between EUV and X-ray flux variations from \ngc\ in simultaneous observations
with \asca, \xte\ and \euve\ (Chiang \et\ 2000). 

In principle it should be possible to distinguish between the above
two models in the higher energy spectral band where the ionised
reflector will continue to make a contribution. We have attempted this
comparison by extending the Comptonised thermal and ionised reflector
fits to 100 keV by inclusion of the PDS data. The power 
law index was fixed in each case at the MOS2 fit 
value of $\Gamma$=1.75, and the warm absorber and Fe K emission line 
parameters were also retained from the 0.3-10 keV fits. Good fits were
obtained with both models, which yielded interesting (but not decisive)
differences in 10-100 keV band.
The Comptonised thermal description of the soft excess ($\chi^{2}$/dof
= 576/590)
has essentially
no effect in the higher energy band (Figure 9), with all the continuum
'hump' being due to the cold reflector. In the case of the ionised
reflector fit (only marginally worse with $\chi^{2}$/dof = 593/590),
there is a contribution across the whole observed band (Figure 10). 
In consequence,
the Fe 
abundance in PEXRAV drops from $\sim$1.3 to $\sim$0.3 to compensate for 
the 'filling in' of the observed flux at $\sim$7-15 keV.
In addition the high energy cut-off energy of $\sim$150 keV in the
Comptonisation fit extends beyond $\sim$300 keV in the ionised disc fit.   
We conclude that, in the present case, our high energy data are unable 
to distinguish between the two
models for the (relatively weak) soft excess.

\begin{figure}
\centering
\includegraphics[width=6 cm, angle=270]{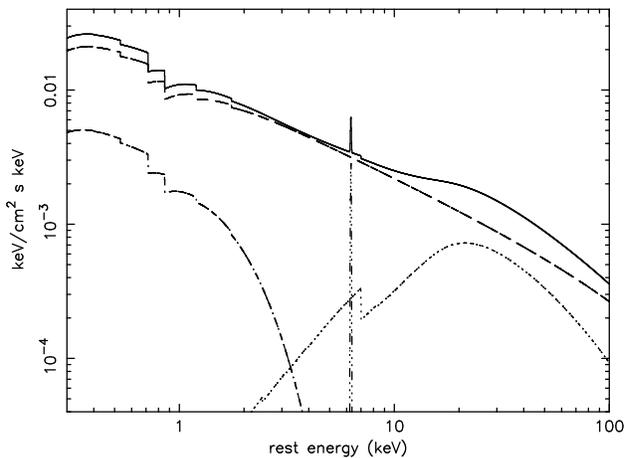}
\caption
{Power law, cold reflection and Comptonised thermal emission model fit
to the MOS2 and PDS data.
See text for further
details}
\end{figure}

\begin{figure}
\centering
\includegraphics[width=6 cm, angle=270]{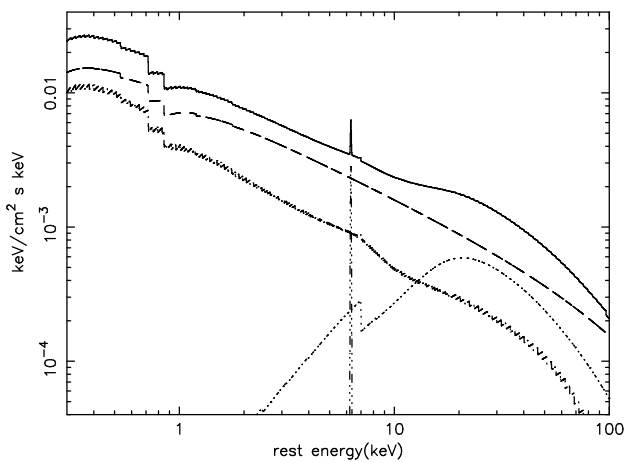}
\caption
{Power law, cold reflection and ionised reflection fit to the MOS2 and
PDS data. The latter
is seen to make only a small contribution to the high
energy `hump'. See text for further details.}
\end{figure}

\subsection{The X-ray signature of \ngc, a low accretion rate AGN}

The average observed flux in the 2--10~keV band during orbit 290 
was $5 \times       
10^{-11}$~erg s$^{-1}$  cm$^{-2}$, corresponding to a    
luminosity of $\sim 2.5 \times 10^{43}$~erg s$^{-1}$. The bolometric   
luminosity is then likely to be of order $\sim 5 \times 10^{44}$~erg  
s$^{-1}$. Assuming a black 
hole mass for \ngc\ of $M \sim 6 \times 10^{7}\Msun$ (Ferrarese \et\
2001) this luminosity requires
an accretion rate of $\sim$ 4 percent of the Eddington rate.
We suggest this may be a typical value for many Broad Line
Seyfert 1 galaxies.

The corresponding X-ray signature of Broad Line Seyfert 1s might then be much as we find for
\ngc, namely a dominant power law component with (hard) photon index $\sim$
1.75, a weak but relatively `hot' soft excess, and a narrow, `cold' Fe
K emission line. The most clear cut reflection component, dominating
the $\sim$ 10-100 keV spectrum of \ngc, appears also to be from cold (perhaps
the same) matter subtending $\sim$1-1.5$\pi$ steradian. The emergence
of a similar value in Compton-thick Seyfert 2 galaxies
(Matt 2002, Schurch 2002) suggests that substantial reflection from cold 
matter, distant from the 
hard X-ray
source, is a common property of all Seyfert galaxies. If so, this
underlines the importance of taking due
account of such reflection in any analysis or modelling of 
AGN spectra.  

The absence of a broad iron K line is consistent with the inner
accretion disc being highly ionised by the relatively hard incident spectrum, or
simply absent at the low implied accretion rate. We note that the former
explanation would be qualitatively consistent with one of the two
physical models fitted to the soft excess, with magnetic flares
providing the localised source(s) of 
hard X-ray flux. 

In the context of the alternative, thermal Comptonisation description
of the soft excess, it is interesting  to recall that \exosat\ data
showed
that the `soft excess' in \ngc\ was only apparent when the
source flux exceeded $3.5 \times                        
10^{-11}$~erg s$^{-1}$  cm$^{-2}$ (Branduardi-Raymont \et\ 1986), as it is 
here.
We note that a flux-related `switching on' of the disc X-ray
emission in AGN could be a further analogy with the X-ray properties 
of Galactic 
black hole candidate sources, which are found to change from a
faint/hard state when the accretion rate rises above $\sim$ a
few
percent of the Eddington rate (eg Chen \et\ 1995). A common physical explanation might then be 
sought in terms of
the well known radiation pressure instability (Shapiro \et\ 1976), 
with re-structuring the inner disc leading to
Comptonisation of the thermal disc photons at a relative
accretion rate of a few percent (Kubota \et\ 2001, Done 2002).

A truncated disc provides the most obvious alternative (to the highly 
ionised inner
disc), consistent with the X-ray signature of \ngc, namely hard power
law spectrum,
moderate luminosity, weak soft excess, and absence of a broad iron K
line. In this scenario the inner radius of the dense matter should penetrate 
further into a hot inner flow as the 
accretion
rate increased, perhaps giving a more 
gradual spectral change.
 
Future studies of the broad band spectral variability of sources such
as \ngc\ will be important to search for such effects, as will the testing of
models over a wider sample of AGN, with a larger range of spectral
properties. 

\section{Conclusions}

(1) A long \xmm\ and \sax\ observation of the bright broad line Seyfert 1
galaxy finds a broad-band X-ray spectrum dominated by a hard power
law of index $\Gamma$$\sim$1.75, a value derived after due allowance
for a continuum reflection component peaking at $\sim$30-40keV.

(2) The continuum reflection corresponds to cold (neutral)
matter subtending a solid angle of $\sim$1-1.5$\pi$ steradian at
the hard X-ray source. We suggest this may be a very common component in 
the nuclei of Seyfert galaxies.

(3) The only spectral feature seen in the 3-10 keV spectrum is a
narrow emission line at $\sim$6.4 keV with an equivalent width of
$\sim$60 eV, consistent with iron fluorescence from the same cold
matter. The lack of variability over a 17 month interval between
\chandra\ and \xmm\ observations is consistent with this matter being
in the putative torus; however, the marginally resolved line width
suggests a significant component lies at a smaller radius.

(4) A weak soft excess is seen, after allowing for over-lying
absorption, as a smooth upward curvature in the X-ray
continuum below $\sim$2 keV, and can be modelled by 2 blackbodies,
Comptonised thermal emission from the accretion disc, or ionised
disc reflection.   

(5) The overall X-ray emission spectrum is consistent with a model in which
accretion at a relatively low rate (a few percent of
the Eddington rate) is depositing most of the corresponding accretion energy
in a hot, low density medium (or corona), with the dense inner disc    
being absent (a truncated disc) or highly ionised.

(6) The 3 $\sigma$ upper limit of 43 eV on a broad iron K line confirms the
inner disc to be in a very different state from that indicated by the earlier
\asca\ observation, although the hard X-ray power law emission was
essentially the same.

\section*{ Acknowledgements }
The results reported here are based primarily on observations obtained with \xmm, an ESA science mission with
instruments and contributions directly funded by ESA Member States and
the USA (NASA).
The authors wish to thank the SOC and SSC teams for organising the \xmm\
observations and initial data reduction. RE acknowledges support from the NASA
XMM-Newton grant NAG 5-10032. JNR and KLP are supported by a Leverhulme
Fellowship and PPARC studentship, respectively. GM and GCP acknowledge
support from ASI and MIUR under the grant \sc cofin-00-02-36.

\end{document}